\documentclass[12pt]{iopart}

\usepackage{braket}
\usepackage{color}
\usepackage{amsmath}
\usepackage{subcaption}
\usepackage{graphicx}
\usepackage{hyperref}
\usepackage{comment}

\begin{document}
\title{Manipulating the Dipolar Interactions and Cooperative Effects in Confined Geometries}

\author{Hadiseh Alaeian}
\address{Elmore Family School of Electrical and Computer Engineering, Department of Physics and Astronomy, Purdue Quantum Science and Engineering Institute, Purdue University, West Lafayette, IN 47907, US}

\author{Artur Skljarow}
\address{5 Physikalisches Institut, Center for Integrated Quantum Science and Technology (IQST),
Universit\"at Stuttgart, Pfaffenwaldring 57, 70569 Stuttgart, Germany}

\author{Stefan Scheel}
\address{Institut f\"ur Physik, Universit\"at Rostock, Albert-Einstein-Stra{\ss}e 23-24, 18059 Rostock, Germany}

\author{Tilman Pfau}
\address{5 Physikalisches Institut, Center for Integrated Quantum Science and Technology (IQST),
Universit\"at Stuttgart, Pfaffenwaldring 57, 70569 Stuttgart, Germany}

\author{Robert L\"ow}
\address{5 Physikalisches Institut, Center for Integrated Quantum Science and Technology (IQST),
Universit\"at Stuttgart, Pfaffenwaldring 57, 70569 Stuttgart, Germany}

\ead{halaeian@purdue.edu}
\vspace{10pt}
\begin{indented}
\item[\today]
\end{indented}

\begin{abstract}
To facilitate the transition of quantum effects from the controlled laboratory environment to practical real-world applications, there is a pressing need for scalable platforms. One promising strategy involves integrating thermal vapors with nanostructures designed to manipulate atomic interactions. In this tutorial, we aim to gain deeper insights into this by examining the behavior of thermal vapors that are confined within nanocavities or waveguides and exposed to near-resonant light.
We explore the interactions between atoms in confined dense thermal vapors. Our investigation reveals deviations from the predictions of continuous electrodynamics models, including density-dependent line shifts and broadening effects. In particular, our results demonstrate that by carefully controlling the saturation of single atoms and the interactions among multiple atoms using nanostructures, along with controlling the geometry of the atomic cloud, it becomes possible to manipulate the effective optical nonlinearity of the entire atomic ensemble. This capability renders the hybrid thermal atom-nanophotonic platform a distinctive and valuable one for manipulating the collective effect and achieving substantial optical nonlinearities.
\end{abstract}

%

\section{Introduction}~\label{sec: introduction}
Atoms represent ideal systems for investigating cooperative quantum phenomena due to their narrow radiative linewidth and consistent reproducibility. When photons are emitted by two distinct atoms, they often exhibit remarkable similarity, setting atoms apart from artificial light emitters such as quantum dots, vacancy centers, or rare-earth ions. Unlike these artificial emitters, atoms cannot be customized to user specifications but offer predictable properties with additional interactions involving their environment, like phonons or nuclear spins.

In certain situations, cold and thermal atomic clouds can be confined very close to macroscopic surfaces, such as wedged vapor cells~\cite{Ripka2018, Christaller2022}, nano-fibers~\cite{Mitsch2014, Kien2017}, or atom-cladded nanophotonic devices~\cite{Thompson2013, Stern2013, Goban2015, Ritter2015, Ritter2016, Stern2017, Ritter2018, Skljarow2020, Skljarow2022}. In these cases, the electromagnetic characteristics of the macroscopic surroundings influence the interactions between the atomic dipoles. These effects can be analyzed within the framework of macroscopic quantum electrodynamics (QED)~\cite{Sandoghdar1992, Scheel2008}, which describes the quantized electromagnetic field in various absorptive and dispersive linear media using the classical electromagnetic Green’s tensor. This can be seen as a generalized mode expansion where photons occupy the same modes as classical wave counterparts. The theory accounts for quantum fluctuations in matter and fields, which can spontaneously polarize neutral atoms, leading to effective interactions with the surrounding macroscopic objects. This interaction with surfaces results in an additional radiative line shift, known as the Casimir-Polder shift~\cite{Casimir1948}. It is one of several electromagnetic dispersion interactions caused by quantum vacuum fluctuations, depending on variations in atom and body properties with frequency.

Furthermore, in an ensemble of atoms where many dipoles interact simultaneously, new properties can emerge. Notable examples are altered spontaneous emission rates, referred to as sub- or superradiance, initially predicted by Dicke in 1954~\cite{Dicke1954} and elaborated on by Lehmberg~\cite{Lehmberg70a, Lehmberg70b}. Resonant atom-atom interactions can have both positive and negative effects. For instance, in an array of atomic clocks, undesired interactions reduce coherence and precision. Conversely, they can be harnessed to engineer light-matter interactions, creating a perfect mirror using a single atomic or excitonic layer~\cite{Bettles2016, Shahmoon2017, Zeytino2017, Back2018, Rui2020}.

In a continuous medium, a polarized atom alters its local environment, affecting the local field at the position of other atoms, known as the local field correction. This is described by the density-dependent Lorentz–Lorenz shift. In a discrete medium of closely spaced individual atoms, superradiance results in a dispersive effective interaction, sometimes referred to as the cooperative Lamb shift~\cite{Friedberg72}, observed in thin atomic layers~\cite{Keaveney12, Peyrot18, Peyrot19}. Various theoretical approaches have been proposed~\cite{Ruostekoski1997, Javanainen14, Dobbertin2020} to explain some but not all observed features. In systems with numerous interacting atoms in a disordered ensemble, the resulting inhomogeneous broadening of atomic resonance frequencies leads to saturation of the optical response, resulting in an effective refractive index of the atomic medium that reaches approximately $n\approx1.7$~\cite{Andreoli21}.

This work employs a first-principle approach to unify interpretations and treatments of dipole-dipole interactions (DDI) and cooperative effects in the literature. While the primary focus is on thermal atomic ensembles experiencing motional dephasing, the theoretical framework is broadly applicable. The paper's structure is as follows: Section~\ref{sec: DDI derivation} uses the most general formalism of macroscopic quantum electrodynamics to derive the dipole-dipole interaction, starting from the general quantum master equation. It discusses the impact of resonant and virtual photon transitions on cooperative shifts and interactions, shedding light on different approximations commonly used in literature to describe cooperative effects in various physical platforms across different limits. To the best of our knowledge, the approach presented here represents the first comprehensive quantum treatment of this problem. The section concludes by deriving the well-known spin model often used to describe cooperative effects in atomic systems.

In Section~\ref{sec: interacting cloud}, the spin model is employed to calculate the evolution of the joint density matrix of the atomic ensemble in free space. For the two-atom case, the effective non-Hermitian Hamiltonian is directly diagonalized to determine the energies and decay rates of hybridized states, illustrating their sub(super)-radiant characteristics. The study is then extended to larger ensembles with more than two atoms, with a focus on thermal atoms subject to transit and Doppler broadening and dephasing. While exact diagonalization of the ensemble effective Hamiltonian becomes impractical in this limit due to the rapid dephasing of thermal atoms, rendering the mean-field (MF) approximation a valid approach. A Monte-Carlo approach is used to account for all motional effects, examining dipolar interactions in various dimensions, from 1D to 3D clouds, emphasizing the crucial role of geometry and dimensionality in emerging dipolar effects.

Section~\ref{sec: experiment} summarizes recent experiments that utilize thermal atoms to investigate and manipulate dipolar interactions. Finally, Section~\ref{sec: conclusion} provides a summary of the results and suggests intriguing directions for exploring dipolar atomic ensembles at various dimensions.

\section{Atom-atom interactions from first principles}~\label{sec: DDI derivation}
The interaction of an atomic ensemble with a common light field leads to an effective light-
induced dipole-dipole interaction as well as cooperative effects~\cite{Passante2018, Reitz2022}. 
To derive the simplest case of light-induced dipolar interaction from first principles, we consider $N$ atoms interacting with the electromagnetic field in free space. After tracing over the electromagnetic degrees of freedom, one obtains an effective dipole-dipole interaction between the atoms. In what follows, we will start from the most general description of an atomic 
ensemble interacting with quantized electromagnetic fields to derive a quantum master equation that describes the evolution of the reduced atomic density matrix. Then, we show how, by employing several approximations such as the rotating wave, Born-Markov, and mean-field 
approximations, one obtains various descriptions of the interacting emitter ensemble including the well-known classical coupled dipole model. This pedagogical approach presented here clarifies the 
range of validity of each approximation and delimits their applicability in modeling the 
experimental scenarios and describing their results as summarized in Sec.~\ref{sec: experiment}.

In the most general form, and when expanded in Fourier space, the electric field operator 
reads~\cite{Scheel2008}
\begin{equation}~\label{eq:photon field operator}
    \hat{\mathbf{E}}(\mathbf{R}) = \int_0^\infty d\omega \left(\hat{\mathbf{\mathcal{E}}}(\mathbf{R},\omega) + \hat{\mathbf{\mathcal{E}}}^\dagger(\mathbf{R},\omega) \right)\, ,
\end{equation}
where $\hat{\mathbf{\mathcal{E}}}(\mathbf{R},\omega)$ is related to $\hat{\mathbf{f}}(\mathbf{R},\omega)$, 
the annihilation operator of the medium-assisted field excitation at position $\mathbf{R}$ and 
frequency $\omega$, as~\cite{Perina2001, Scheel2008}
\begin{equation}~\label{eq:quadrature via annihilation}
\hat{\mathbf{\mathcal{E}}}(\mathbf{R},\omega) = i \left(\frac{\omega     }{c}\right)^2 \sqrt{\frac{\hbar}{\pi \epsilon_0}} \int d\mathbf{R'} \sqrt{\textrm{Im}(\epsilon(\mathbf{R'},\omega))} \,G(\mathbf{R}, \mathbf{R'}, \omega)~\hat{\mathbf{f}}(\mathbf{R'},\omega)\, ,
\end{equation}
where $\epsilon(\mathbf{R},\omega)$ is the permittivity of the medium and 
$G(\mathbf{R}, \mathbf{R'}, \omega)$ is the classical electromagnetic Green's tensor determined via the Helmholtz equation~\cite{Jackson1999}
\begin{equation}
    \left[\nabla \times \nabla \times - \left(\frac{\omega}{c}\right)^2 \epsilon(\mathbf{R},\omega) \right]G(\mathbf{R}, \mathbf{R'}, \omega) = \delta(\mathbf{R} - \mathbf{R'}) I\, .
\end{equation}

For an ensemble of $N$ multi-level atoms with transition frequencies $\omega_n$ interacting with 
the continuum modes of the electromagnetic field, the total Hamiltonian reads as~\cite{Cohen-Tannoudji1973}
\begin{equation}~\label{eq:total H}
    \hat{\mathcal{H}} = \hat{H}_A + \hat{H}_F + \hat{H}_{AF}= \hbar \sum_{A,n} \omega_{A,n} 
    \hat{\sigma}_{A,nn} + \hbar \int d\mathbf{R} \int_0^\infty d\omega ~ \omega\, 
    \hat{\mathbf{f}}^\dagger \hat{\mathbf{f}} - \sum_{A} \hat{\mathbf{d}}_A \cdot 
    \hat{\mathbf{E}}(\mathbf{R}_A)  \, ,
\end{equation}
where the first and second terms correspond to the free atom and the electromagnetic field energies, respectively, and the last term describes the atom-field interaction to the leading order, i.e. in electric-dipole approximation.

Further, $\hat{\sigma}_{A,mn} = \ket{m_A}\bra{n_A}$ are the spin transition operators between electronic levels $n,m$ of the $A^\textrm{th}$ atom, and $\hat{\mathbf{d}}_A$ is the transition dipole moment operator of the $A^\textrm{th}$ atom as, 
\begin{equation}~\label{eq:dipole moment}
    \hat{\mathbf{d}}_A = -q \sum_{m,n} \braket{m_A|\mathbf{r}_A|n_A} \hat{\sigma}_{A,mn}  
    = \mathbf{d}_{A,mn} \hat{\sigma}_{A,mn}\, .
\end{equation}
Note that, for a two-level atom, the above dipole operator simplifies to 
$\mathbf{d} \left(\hat{\sigma}_A^+ + \hat{\sigma}_A^- \right)$ in terms of the well-known raising 
and lowering spin operators $\hat{\sigma}_A^\pm$. Although a real atom is not a simple two-level system, later it will become evident under which circumstances this is a 
sensible approximation.

The joint atom-field density operator evolves according to the von Neumann equation as
\begin{equation}~\label{eq:von Neumann}
    \dot{\hat{\rho}}(t) = - \frac{i}{\hbar} \left[\hat{H}_A + \hat{H}_F + \hat{H}_{AF} , \hat{\rho}(t)\right]\, ,
\end{equation}
with the formal solution, valid for a time-independent Hamiltonian,
\begin{equation}~\label{eq:formal density solution}
    \hat{\rho}(t) = \hat{U}(t) \hat{\rho}(0) \hat{U}^\dagger(t) ~, ~ 
    \hat{U}(t) = e^{-i t \hat{\mathcal{H}}/\hbar} \,.
\end{equation}
To determine the atom-atom interaction, the electromagnetic degrees of freedom have to 
be integrated out to obtain the dynamics in terms of atomic quantities, only. This will become 
clearer in the interaction picture after rotating the joint atom-field density operator to the 
frame of the free atoms and photons given by $\hat{\mathcal{H}}_0 = \hat{H}_A + \hat{H}_F$ via 
a unitary transformation $\hat{U}_0(t) = e^{-i t \hat{\mathcal{H}}_0/\hbar}$ 
\begin{equation}~\label{eq:interaction operator}
    \hat{\mathcal{O}}_I(t) = \hat{U}_0^\dagger(t) \hat{\mathcal{O}}(t)  \hat{U}_0(t) \,.
\end{equation}
It is straightforward to show that the time evolution of the rotated joint density matrix in the interaction picture is given by
\begin{equation}~\label{eq:interaction density matrix evolution}
    \dot{\hat{\rho}}_I(t) = - \frac{i}{\hbar} \left[\hat{H}_{AF_I} (t) , \hat{\rho}_I(t)\right] \, .
\end{equation}
The reduced atomic density matrix can be obtained by taking the partial trace over the 
electromagnetic field degrees of freedom as
\begin{equation}~\label{eq:atomic reduced density matrix}
    \hat{\rho}_I^A(t) = \textrm{Tr}_F\left(\hat{\rho}_I(t) \right)\, .
\end{equation}
As the goal is to derive an explicit equation of motion for the reduced atomic density matrix, we can directly use Eq.~(\ref{eq:atomic reduced density matrix}) in 
Eq.~(\ref{eq:interaction density matrix evolution}) to find the time evolution of 
$\hat{\rho}_I^A(t)$. 

At this point, it is helpful to separate the interaction of the atom with the electromagnetic field into two parts; 1) the interaction with the driving field which is typically in a coherent 
state, and 2) the interaction with other allowed modes, e.g. via the incoherent spontaneous 
emission into other electromagnetic modes in free space. Here the subscripts 
$L$ and $V$ refer to the first and the second case, respectively. We formally write the time 
evolution of the reduced atomic density matrix as
\begin{equation}~\label{eq:atomic reduced density matrix2}
    \dot{\hat{\rho}}_I^A(t) = - \frac{i}{\hbar} \textrm{Tr}_L \left[\hat{H}_{AF_I}(t) , \hat{\rho}_I(t) \right] - \frac{i}{\hbar} \textrm{Tr}_V \left[\hat{H}_{AF_I}(t) , \hat{\rho}_I(t) \right] \, .
\end{equation}

The incoherently populated fluctuating electromagnetic modes can be described as spatially and spectrally 
uncorrelated fields with the following properties~\cite{Buhmann2008}:
\begin{eqnarray}~\label{eq:field noise correlations}
    \braket{\hat{\mathbf{E}}(\mathbf{R},\omega)}_V & = & 0 \,, \nonumber \\
    \braket{\hat{\mathbf{E}}(\mathbf{R},\omega) \otimes \hat{\mathbf{E}}(\mathbf{R'},\omega')}_V & =& 0 \, ,  \\    \braket{\hat{\mathbf{E}}^\dagger(\mathbf{R},\omega) \otimes \hat{\mathbf{E}}(\mathbf{R'},\omega')}_V & =& \frac{\hbar \mu_0 \omega^2}{\pi} n_T(\omega) \textrm{Im}(G(\mathbf{R},\mathbf{R'},\omega)) \delta(\omega - \omega')\, , \nonumber
    \\
    \braket{\hat{\mathbf{E}}(\mathbf{R},\omega) \otimes \hat{\mathbf{E}}^\dagger(\mathbf{R'},\omega')}_V & = &\frac{\hbar \mu_0 \omega^2}{\pi}\left(1 + n_T(\omega) \right) \textrm{Im}(G(\mathbf{R},\mathbf{R'},\omega)) \delta(\omega - \omega')\, , \nonumber
\end{eqnarray}
where $n_T(\omega)$ is the mean number of thermal photons at temperature $T$ and frequency $\omega$ determined via the Bose-Einstein distribution as
\begin{equation}~\label{eq:BE distribution}
    n_T(\omega) = \frac{1}{e^{\frac{\hbar\omega}{k_B T}} - 1} \, .
\end{equation}
As can be seen, the electric field vanishes on average, consistent with the assumption about their population via fluctuations, but the field intensity does not. Later, it will become clear that these contributions lead to stimulated emission ($\propto n_T(\omega)$) and absorption ($\propto 1 + n_T(\omega)$) triggered by thermal photons and the vacuum. The latter case is better known as spontaneous emission.

Equation~(\ref{eq:interaction density matrix evolution}) can be formally solved for the 
interacting density matrix as
\begin{equation}~\label{eq:formal solution of interaction density}
    \hat{\rho}_I(t) = \hat{\rho}_I(0) - \frac{i}{\hbar} \int_0^t dt' ~ \left[\hat{H}_{AF_I}(t') , \hat{\rho}_I(t') \right] \, .
\end{equation}
As can be seen from Eq.~(\ref{eq:field noise correlations}), the non-driven, spontaneously 
occupied modes are described via their correlations, which implies that the formal solution
Eq.~(\ref{eq:formal solution of interaction density}) should be re-inserted into the second term 
of Eq.~(\ref{eq:atomic reduced density matrix2}), i.e. fluctuating field contributions, leading to the equation of motion for the 
reduced atomic density matrix as
\begin{equation}~\label{eq:reduced density}
\dot{\hat{\rho}}_I^A(t) = - \frac{i}{\hbar} \textrm{Tr}_L \left[\hat{H}_{AF_I}(t) , \hat{\rho}_I(t) \right] - \frac{i}{\hbar} \textrm{Tr}_V \left[\hat{H}_{AF_I}(t) , \hat{\rho}_I(0) \right] - \frac{1}{\hbar^2}\textrm{Tr}_V \int_0^t dt'~ \left[\hat{H}_{AF_I}(t) , \left[\hat{H}_{AF_I}(t') , \hat{\rho}_I(t') \right]\right]\, .
\end{equation}

We assume that, at $t=0$, atoms and fields are not correlated, and hence 
$\hat{\rho}_I(0) = \hat{\rho}_A(0) \otimes \hat{\rho}_F$~\footnote{This is a strong assumption that has to be checked in each scenario.}. 
Under this condition the second term in 
Eq.~(\ref{eq:reduced density}) vanishes. 

Under the weak atom-field couplings, the field remains unchanged, i.e. $\dot{\hat{\rho}}_F(t) = 0$. This limit, known as \emph{Born} 
approximation allows one to write the joint density matrix as a direct product of atoms and 
photon density matrices at all times, i.e. $\hat{\rho}_I(t) \approx \hat{\rho}_I^A(t) \otimes \hat{\rho}_F$~\cite{Breuer2002}. In particular, this means that the joint evolution does not correlate or entangle atoms and light at 
any time step.

Furthermore, if the atomic states change slowly enough compared to all other timescales, their 
time evolution is a \emph{Markovian} process implying that 
$\hat{\rho}_I^A(t') \approx\hat{\rho}_I^A(t)$~\footnote{This approximation is justified since the reduced density is already in the rotated frame and hence the fast-varying part has been already extracted.}. With these assumptions, Eq.~(\ref{eq:reduced density}) can be simplified to~\cite{Gardiner2004} 
\begin{equation}~\label{eq:reduced density simplified}
\dot{\hat{\rho}}_I^A(t) = - \frac{i}{\hbar} \textrm{Tr}_L \left[\hat{H}_{AF_I}(t) , \hat{\rho}_I^A(t) \otimes \hat{\rho}_F \right] - \frac{1}{\hbar^2}\textrm{Tr}_V \int\limits_0^\infty dt'~ \left[\hat{H}_{AF_I}(t) , \left[\hat{H}_{AF_I}(t - t') , \hat{\rho}_I^A(t) \otimes \hat{\rho}_F \right]\right],
\end{equation}
where in the last integral the limit is extended to $\infty$, justified by the finite correlation times of the field. 

\subsection{Quantum master equation for the reduced atomic density matrix}~\label{subsec: interaction frame}
Starting from Eq.~(\ref{eq:reduced density simplified}), we first calculate the rotated interacting 
Hamiltonian $\hat{H}_{AF_I}$ as
\begin{equation}
    \hat{H}_{AF_I}(t) =  - \sum_{B} \left(e^{it \hat{H}_A/\hbar} \hat{\mathbf{d}}_B e^{-it \hat{H}_A/\hbar} \right) \cdot \left(e^{it \hat{H}_F/\hbar} \hat{\mathbf{E}}(\mathbf{R}_B)  e^{-it \hat{H}_F/\hbar}\right)\, ,
\end{equation}
where we used $\left[\hat{H}_A , \hat{H}_F\right] = 0$, to separate the unitary rotation $\hat{U}_0(t)$ into atom and field parts.

Using the explicit form of the atom-field interaction in the electric-dipole approximation limit, the Hamiltonian in the interaction picture reads as 
\begin{equation}~\label{eq:rotated interaction Hamiltonian}
   \hat{H}_{AF_I}= - \sum_{B} \underbrace{\sum_{m,n} \mathbf{d}_{B,mn} e^{i\omega_{B,mn} t} \hat{\sigma}_{B,mn} }_{\hat{\mathbf{d}}_{B_I}(t)} \cdot \underbrace{\int_0^\infty d\omega ~ \left(e^{i\omega t} \hat{\mathcal{E}}^\dagger(\mathbf{R}_B,\omega) + e^{-i\omega t} \hat{\mathcal{E}}(\mathbf{R}_B,\omega) \right)}_{\hat{H}_{F_I}(t)}  \, .
\end{equation}
Since the interaction Hamiltonian can be decomposed as 
$\hat{H}_{AF_I} = \hat{H}_{A_I} \otimes \hat{H}_{F_I}$, 
one can directly calculate the contribution of the driving field in 
Eq.~(\ref{eq:reduced density simplified}) as \cite{Altafini2004}
\begin{eqnarray}
    \textrm{T}_L &=&   - \frac{i}{\hbar} \textrm{Tr}_L \left[\hat{H}_{AF_I}(t) , \hat{\rho}_I^A(t) \otimes \hat{\rho}_F \right] 
    \nonumber\\
    &=& -\frac{i}{2\hbar} \textrm{Tr}_L  \Big( \left[ \hat{H}_{A_I}(t) ,  \hat{\rho}_I^A(t) \right] 
    \otimes \left\{ \hat{H}_{F_I} , \hat{\rho}_F \right\} 
    \nonumber \\ && + \left\{ \hat{H}_{A_I}(t) ,  \hat{\rho}_I^A(t) \right\} 
    \otimes \left[ \hat{H}_{F_I} , \hat{\rho}_F \right] \Big) 
    \nonumber\\
    &=& -\frac{i}{\hbar} \textrm{Tr}_L \left( \hat{H}_{F_I} \hat{\rho}_F \right) \left[ \hat{H}_{A_I}(t) ,  \hat{\rho}_I^A(t) \right] \, ,
\end{eqnarray}
where we used $\textrm{Tr}_L[\hat{H}_{F_I} , \hat{\rho}_F]=0$ and 
$\textrm{Tr}_L\{\hat{H}_{F_I} , \hat{\rho}_F\}=2\textrm{Tr}_L( \hat{H}_{F_I} \hat{\rho}_F)$.

Assuming the typical spectroscopy scenario using a coherent field $\ket{E_{\mathbf{r},\omega}}$, the field density 
operator can be written as $\hat{\rho}_F = \ket{E_{\mathbf{r},\omega}}\bra{E_{\mathbf{r},\omega}}$ and hence the partial trace $\textrm{Tr}_L$ over the field can be simplified as
\begin{eqnarray}~\label{eq:Tl}
    \textrm{Tr}_L \left(\hat{H}_{F_I}  \hat{\rho}_F\right) 
   & = \textrm{Tr}_L \int_0^\infty d\omega  \left(
 e^{i\omega t} \hat{\mathcal{E}}^\dagger(\mathbf{R}_B,\omega) + e^{-i\omega t} \hat{\mathcal{E}}(\mathbf{R}_B,\omega)   
    \right) \ket{E_{\mathbf{r},\omega}}\bra{E_{\mathbf{r},\omega}}  \nonumber \\
   & = \int_0^\infty d\omega ~ \left(e^{i\omega t} \mathbf{E}_{\mathbf{r},\omega}^* +  e^{-i\omega t} 
    \mathbf{E}_{\mathbf{r},\omega} \right)  
    = \mathbf{E}(\mathbf{r},t) \, ,
\end{eqnarray}
where in the last line we employed the unity trace property of the density matrix $\hat{\rho}_F$ followed by the inverse Fourier transform to re-constitute the driving field in the time
domain from its frequency components. 

This simplifies the driving field contribution to the 
evolution of the reduced atomic density matrix to
\begin{equation}~\label{eq:drive field contribution}
    \textrm{T}_L = \frac{i}{\hbar} \left[\sum_{B} \hat{\mathbf{d}}_{B_I}(t) \cdot \mathbf{E}(\mathbf{R}_B,t) , \hat{\rho}_I^A(t) \right]  \, .
\end{equation}
The final equation is the Rabi Hamiltonian describing the dynamics of atoms driven by the classical field 
$\mathbf{E}(\mathbf{R}_B,t)$. As can be seen, the coupling of each atom to the driving field is 
independent of other atoms in the ensemble which is consistent with the aforementioned Born 
approximation. 

On the other hand, there will be effective atom-atom interactions that arise from the second term in 
Eq.~(\ref{eq:reduced density simplified}). Those modes are not externally driven and are 
solely populated by photons emitted by the atoms, whose contributions can be determined as
\begin{eqnarray}~\label{eq:Tv}
    \textrm{T}_V  = & - \frac{1}{\hbar^2} \textrm{Tr}_V \int_0^\infty dt'~ \left[\hat{H}_{AF_I}(t) , \left[\hat{H}_{AF_I}(t - t') , \hat{\rho}_I^A(t) \otimes \hat{\rho}_F \right]\right] \nonumber \\
      = & - \frac{1}{\hbar^2}\sum_{C,D} \int_0^\infty dt' ~ \hat{\mathbf{d}}_{C_I}(t) \cdot \textrm{Tr}_V \{\hat{\mathbf{E}}_I(\mathbf{R}_C,t) \otimes \hat{\mathbf{E}}_I(\mathbf{R}_D,t - t') \hat{\rho}_{F_I}\} \hat{\mathbf{d}}_{D_I}(t - t') \hat{\rho}_I^A(t) \nonumber \\
     &  + \frac{1}{\hbar^2}\sum_{C,D} \int_0^\infty dt' ~ \hat{\mathbf{d}}_{C_I}(t) \hat{\rho}_I^A(t) \cdot \textrm{Tr}_V \{\hat{\mathbf{E}}_I(\mathbf{R}_C,t) \otimes \hat{\mathbf{E}}_I(\mathbf{R}_D,t - t') \hat{\rho}_{F_I}\} \hat{\mathbf{d}}_{D_I}(t - t')  \nonumber \\
    &  + \frac{1}{\hbar^2} \sum_{C,D} \int_0^\infty dt' ~ \hat{\mathbf{d}}_{D_I}(t - t') \cdot \textrm{Tr}_V \{\hat{\mathbf{E}}_I(\mathbf{R}_D,t - t') \otimes \hat{\mathbf{E}}_I(\mathbf{R}_C,t) \hat{\rho}_{F_I}\} \hat{\rho}_I^A(t) \hat{\mathbf{d}}_{C_I}(t)  \nonumber \\
    & - \frac{1}{\hbar^2} \sum_{C,D} \int_0^\infty dt' \hat{\rho}_I^A(t) ~ \hat{\mathbf{d}}_{D_I}(t - t')  \cdot \textrm{Tr}_V \{\hat{\mathbf{E}}_I(\mathbf{R}_D,t - t') \otimes \hat{\mathbf{E}}_I(\mathbf{R}_C,t) \hat{\rho}_{F_I}\} \hat{\mathbf{d}}_{C_I}(t) \, , \nonumber \\
\end{eqnarray}
where we used the identity 
\begin{equation}
    \left(\mathbf{v}_1 \cdot \mathbf{v}_2 \right) \left(\mathbf{v}_3 \cdot \mathbf{v}_4 \right) = \mathbf{v}_1 \cdot \left(\mathbf{v}_2 \otimes \mathbf{v}_3 \right) \mathbf{v}_4
\end{equation}
to re-write the integral kernels as tensor products. 

The correlation properties of these fluctuation fields summarized in Eq.~\ref{eq:field noise correlations} should be used to simplify the integral kernels, after employing the Fourier transformation, as 
\begin{eqnarray}
    \textrm{Tr}_V \{\hat{\mathbf{E}}_I(\mathbf{R}_C,t) \otimes \hat{\mathbf{E}}_I(\mathbf{R}_D,t - t') \hat{\rho}_{F_I}\} \nonumber \\
     = \int_0^\infty d\omega d\omega' \braket{\mathbf{E}(\mathbf{R}_C,\omega)\mathbf{E}^\dagger(\mathbf{R}_D,\omega')}_V e^{-i(\omega - \omega') t - i \omega' t'}  \nonumber \\
     + \int_0^\infty d\omega d\omega' \braket{\mathbf{E}^\dagger(\mathbf{R}_C,\omega)\mathbf{E}(\mathbf{R}_D,\omega')}_V e^{i(\omega - \omega') t + i \omega' t'} \nonumber \\
     = \frac{\hbar \mu_0}{\pi} \int_0^\infty d\omega \left((1 + n(\omega)) e^{-i\omega t'} + n(\omega) e^{i\omega t'}\right) \omega^2 \textrm{Im}(G(\mathbf{R}_C , \mathbf{R}_D,\omega))\, .
\end{eqnarray}

Other terms of Eq.~(\ref{eq:Tv}) can be calculated similarly and expressed in terms of photon density and Green's tensor. 

Combining the results with Eq.~(\ref{eq:drive field contribution}) and inserting the final expression into Eq.~(\ref{eq:reduced density simplified}), we obtain the quantum master equation for the reduced 
atomic density matrix $\hat{\rho}_I^A(t)$ in the interaction picture. The atomic density operator in the original frame can be obtained using the unitary transformation $\hat{\rho}^A(t) = e^{-i t \hat{H}_A/\hbar} \hat{\rho}_I^A(t) e^{i t \hat{H}_A/\hbar}$. Using 
the explicit form of the dipole operator $\mathbf{d}_{C,D}$ in terms of the dipole-allowed transitions as in Eq.~(\ref{eq:dipole moment}), the evolution of the reduced atomic density matrix is given by
\begin{eqnarray}~\label{eq:general reduced density EoM}
    \frac{d}{dt} \hat{\rho}^A(t) = -\frac{i}{\hbar} \left[\hat{H}_A - \sum_{B,m,n} \mathbf{d}_{B,mn} \cdot \mathbf{E}(\mathbf{R}_B , t) \hat{\sigma}_{B,mn}, \hat{\rho}^A(t)\right]  \nonumber \\
    + \sum_{C,D,m,n,i,j} \mathbf{d}_{C,mn} \cdot \mathbf{H}_{ji}(\mathbf{R}_C,\mathbf{R}_D) \cdot \mathbf{d}_{D,ij} \left(\hat{\sigma}_{C,mn} \hat{\sigma}_{D,ij} \hat{\rho}^A(t) - \hat{\sigma}_{D,ij} \hat{\rho}^A(t) \hat{\sigma}_{C,mn}\right) \nonumber \\
    - \sum_{C,D,m,n,i,j} \mathbf{d}_{C,mn} \cdot \mathbf{H}^*_{ji}(\mathbf{R}_C,\mathbf{R}_D) \cdot \mathbf{d}_{D,ij} \left(\hat{\sigma}_{C,mn} \hat{\rho}^A(t) \hat{\sigma}_{D,ij}  - \hat{\rho}^A(t) \hat{\sigma}_{D,ij} \hat{\sigma}_{C,mn}\right) \, ,\nonumber
    \\ 
\end{eqnarray}
where $\mathbf{H}_{ij}(\mathbf{R}_C,\mathbf{R}_D)$ are the Green's function-dependent coefficients for two atoms located at $\mathbf{R}_{C,D}$ for $ji^\textrm
{th}$ transition, as detailed in~\cite{Dobbertin2021}. 

As can be inferred from Eq.~(\ref{eq:general reduced density EoM}), the first line corresponds 
to the single-atom dynamics due to the free atom and the driving field. The second and the third 
lines, on the other hand, describe the light-induced atom-atom interaction, in the forms of spin exchange as 
$\hat{\sigma}_{C,mn} \hat{\sigma}_{D,ij}$. In the next subsection, we show how this general form 
can be used to derive the commonly-used simplified spin model master 
equation~\cite{Manzoni2017, Chang2018, Albrecht2019, Mahmoodian2020, He2020, Moreno-Cardoner2021}
as well as the well-known classical coupled dipole model in some specific limits.

\subsection{The coupled-spin master equation and classical coupled-dipole limit}~\label{subsec: coupled dipolde model}
Using the Heisenberg equation of motion obtained from Eq.~(\ref{eq:general reduced density EoM}), 
one can study the atomic polarization generated by the atom-light interactions as well as the 
light-induced dipole-dipole interaction within the ensemble. Let us assume that the driving field 
is a monochromatic coherent field as $\mathbf{E}_0(\mathbf{r})e^{-i\omega_L t}+\mathrm{c.c.}$, and it is near-resonant with only one of the electronic 
transitions at $\omega_0$, such that $\omega_L - \omega_0 \ll \omega_L - \omega_{mn}$ for $\omega_{mn}$ being the transition frequency of other allowed transitions as ($m \longleftrightarrow n$). This allows one to simplify the equations of motion noticeably, since such near-resonant exciting laser 
couples only those state pairs efficiently, while the far-detuned levels remained almost 
unperturbed and hence can be dropped from Eq.~(\ref{eq:general reduced density EoM}). Further, 
we assume a small detuning as $\omega_L - \omega_0 \ll \omega_L + \omega_0$ which allows one to employ the 
\emph{rotating-wave approximation} to further ignore fast-rotating, i.e. energy non-conserving, 
terms in the Rabi Hamiltonian, and only retain excitation-conserving terms. 

Using common notations of $g,e$ for the states of this two-level atom with transition frequency $\omega_0$ and with sub-levels of $\mu,\nu$, respectively, and in the rotated the frame of the laser $\hat{\tilde{\sigma}}_{B_{\mu \nu}}^{ge}(t) = \hat{\sigma}_{B_{\mu \nu}}^{ge}(t) e^{i\omega_L t}$, the dipole moment of the $A^\textrm{th}$ atom will be given as

\begin{equation}
    \braket{\hat{\mathbf{d}}_{A,ge}} = \Sigma_{\mu , \nu} \braket{\hat{\sigma}_{B_{\mu \nu}}^{ge}} \mathbf{d}_{A_{\mu \nu}}^{ge}\, .
\end{equation}
The time evolution of the spin operator is given via
\begin{eqnarray}~\label{eq:EoM coupled polarization}
    \frac{d}{dt} \braket{\hat{\tilde{\sigma}}_{A_{\mu \nu}}^{ge}(t)} & =  i \left(\omega_L - \omega_0 \right) \braket{\hat{\tilde{\sigma}}_{A_{\mu \nu}}^{ge}(t)}  \\
    & - \frac{i}{\hbar} \sum_\kappa\left(\braket{\hat{\tilde{\sigma}}_{A_{\kappa \nu}}^{ee}(t)} \mathbf{d}_{A_{\kappa \mu}}^{eg} - \braket{\hat{\tilde{\sigma}}_{A_{\mu \kappa}}^{gg}(t)} \mathbf{d}_{A_{\nu \kappa}}^{eg} \right) \cdot \mathbf{E}_0(\mathbf{R}_A) \nonumber \\
    & - i \frac{\mu_0 \omega_0^2}{\hbar}  \sum_{B \ne A} \sum_{\delta \epsilon \kappa}\mathbf{d}_{A_{\kappa \mu}}^{eg} \cdot G(\mathbf{R}_A, \mathbf{R}_B, \omega_0) \cdot \mathbf{d}_{B_{\delta \epsilon}}^{ge} \braket{\hat{\tilde{\sigma}}_{A_{\kappa \nu}}^{ee}(t) \hat{\tilde{\sigma}}_{B_{\delta \epsilon}}^{ge}(t)} \nonumber \\
    & + i \frac{\mu_0 \omega_0^2}{\hbar}  \sum_{B \ne A} \sum_{\delta \epsilon \kappa}\mathbf{d}_{A_{\nu \kappa}}^{eg} \cdot G(\mathbf{R}_A, \mathbf{R}_B, \omega_0) \cdot \mathbf{d}_{B_{\delta \epsilon}}^{ge} \braket{\hat{\tilde{\sigma}}_{A_{\mu \kappa }}^{gg}(t) \hat{\tilde{\sigma}}_{B_{\delta \epsilon}}^{ge}(t)} \nonumber \\
    & + i \sum_\epsilon \left(-\omega_{A_{\nu \epsilon}}^{e^\textrm{CP}}+ i \frac{\Gamma_{A_{\nu \epsilon}}^e}{2}\right) \braket{\hat{\tilde{\sigma}}_{A_{\mu \epsilon}}^{ge}(t)} + \left(\omega_{A_{\mu \epsilon}}^{g^\textrm{CP}}+ i \frac{\Gamma_{A_{\mu \epsilon}}^g}{2}\right)\braket{\hat{\tilde{\sigma}}_{B_{\epsilon \nu}}^{ge}(t)} \nonumber\, .
\end{eqnarray}
This is the \emph{coupled-spin} model, whose first line describes the dynamics of a coherently-driven atom, while the second line is the light-induced dipole-dipole interaction due to the 
exchange of a real photon, i.e. a resonant photon that swaps the spins of the two atoms. 
In the last line $\omega^\textrm{CP}, \Gamma$ corresponds to the collective phase shift and dissipation due to the other off-resonant electronic states of the atom~\cite{Dobbertin2021}. Such a shift was first studied by Casimir and Polder and hence it is preferred 
to as Casimir–Polder (CP) shift~\cite{Casimir1948}, whose detailed form can be found in~\ref{app:CP}.

This treatment also takes into account the modifications of the 
local density of the optical states in any environment if the proper Green's function is used. 
Physically, this leads to a modified decay rate, an effect first predicted and studied by 
Purcell~\cite{Purcell1946, Purcell1995}. 

In general, the coupled-spin master equation of Eq.~(\ref{eq:EoM coupled polarization}) leads to 
an infinite hierarchy of moments for the spins due to the spin correlation terms in the resonant dipole-dipole interaction. However, if such correlations are 
ignored, the \emph{mean-field} approximation can be employed to simplify them as
$\braket{\hat{\sigma}_{C,mn}(t) \hat{\sigma}_{D,ij}(t)} \approx \braket{\hat{\sigma}_{C,mn}(t)} 
\braket{\hat{\sigma}_{D,ij}(t)}$. This approximation converts the equations of motion to a closed system of 
coupled nonlinear equations. Finally, in a weakly-driven limit, where the atoms are not strongly polarized, i.e. $\braket{\hat{\sigma}_A^{ee}(t) - \hat{\sigma}_A^{gg}(t)} \approx -1$, the mean-field version of 
Eq.~(\ref{eq:EoM coupled polarization}) simplifies to the familiar classical coupled-dipole 
model. 

In the next section, we start from the quantum master equation (\ref{eq:EoM coupled polarization}) to 
study the canonical problem of light-induced dipole-dipole interaction between two atoms. 

\section{An interacting ensemble of atoms}~\label{sec: interacting cloud}
Ignoring the fluctuation effects on the lineshift and broadening, and considering a near-resonant monochromatic coherent drive in the form of $\mathbf{E}_0(\mathbf{R})e^{-i
\omega_L t}$ (cf. Sec.~\ref{subsec: coupled dipolde model}), Eq.~(\ref{eq:general reduced density EoM}) for the reduced atomic ensemble density 
operator $\hat{\rho}^A(t)$ can be simplified to the following master equation 
\begin{eqnarray}~\label{eq:QME for atomic density matrix}
    \frac{d}{dt} \hat{\tilde{\rho}}^A(t)  = & -i \left[\sum_{B} \Delta_B \hat{\sigma}_B^+ \hat{\sigma}_B^- - \left(\frac{\mathbf{d}_B \cdot \mathbf{E}_0(\mathbf{R}_B)}{\hbar} \hat{\sigma}_B^+ + \textrm{H.c.}\right) - \sum_{C\ne B} \epsilon_{BC} \hat{\sigma}_B^+ \hat{\sigma}_C^-, \hat{\tilde{\rho}}^A(t)\right]  \nonumber \\
    & + \sum_{B,C} \kappa_{BC} \left(\hat{\sigma}_B^- \hat{\tilde{\rho}}^A(t) \hat{\sigma}_C^+ - \frac{1}{2} \{\hat{\sigma}_B^+ \hat{\sigma}_C^- , \hat{\tilde{\rho}}^A(t)\} \right) \, ,
\end{eqnarray}
where $\Delta_B = \omega_0 - \omega_L$ is the detuning of the $B^\textrm{th}$ atom from the excitation 
laser, $\epsilon_{BC} = 3\pi \Gamma_0 \mathbf{d}_B \cdot \textrm{Re}(G(\mathbf{R}_B,\mathbf{R}_C, 
\omega_0)) \cdot \mathbf{d}_C$ and $\kappa_{BC} = 3\pi \Gamma_0 \mathbf{d}_B \cdot \textrm{Im}
(G(\mathbf{R}_B,\mathbf{R}_C, \omega_0))\cdot\mathbf{d}_C$ correspond to the dispersive and 
dissipative dipole-dipole interactions, respectively. Further, $\Gamma_0$ is the decay rate of a two-level atom in free space and is related to the transition frequency 
$\omega_0$ and dipole matrix element $\mathbf{d}$ as 
\begin{equation}~\label{eq:natural decay rate}
    \Gamma_0 = \frac{\omega_0^3}{3\pi \hbar \epsilon_0 c^3} |\mathbf{d}|^2\, .
\end{equation}
The dyadic Green's function $G(\mathbf{r},\mathbf{r'},\omega)$ depends on the geometry as well as 
the optical properties of the environment, and has a closed-form solution in only a few limited cases. 

In this section, we focus on the physics of such dipolar interactions by limiting ourselves to 
the simplest case, i.e. an atomic ensemble in free space, where a closed-form Green's tensor with the following form is available~\cite{novotny_hecht_2012}
\begin{equation}~\label{eq:vacuum GF}
    G = \left(I + \frac{1}{k_0^2} \nabla \nabla \right) \frac{e^{ik_0 R}}{4\pi k_0 R}\, ,
\end{equation}
where $R = |\mathbf{R}_2 - \mathbf{R}_1|$ and $k_0 = \omega_0/c$.

Using this Green's function, the atom-field interaction energy is given by
\begin{equation}~\label{eq:dispersive DDI}
    \mathcal{G} = \frac{k_0^3 e^{ik_0 R}}{4\pi \epsilon_0 k_0 R} \left[\left(\mathbf{r} \times \mathbf{d}_1 \right) \times \mathbf{r} + \left[3\mathbf{r} \left(\mathbf{r} \cdot \mathbf{d}_1 \right) - \mathbf{d}_1\right] \left(\frac{1}{(k_0 R)^2} - \frac{i}{k_0 R} \right) \right] \cdot \mathbf{d}_2\, ,
\end{equation}
where $\mathbf{r} = (\mathbf{R}_2 - \mathbf{R}_1)/R$ is the unit vector between the two emitters. The dispersive and dissipative interaction effects are related to $\mathcal{G}$ via $\epsilon_{12} = \textrm{Re}(\mathcal{G})$ and $\kappa_{12} = \textrm{Im}(\mathcal{G})$.

\subsection{Canonical problem: two interacting atoms}~\label{subsec: 2-atom case}
%

To develop a better understanding of the dipole-dipole interaction, we
approximate the master equation (\ref{eq:QME for atomic density matrix}) with a non-Hermitian 
Hamiltonian as
\begin{equation}~\label{eq:non-Hermitian H}
    \frac{d}{dt} \hat{\tilde{\rho}}^A(t) = - i\hat{\mathcal{H}}_\textrm{eff} \hat{\tilde{\rho}}^A(t) + i \hat{\tilde{\rho}}^A(t) \hat{\mathcal{H}}_\textrm{eff}^\dagger \,,
\end{equation}
where the effective non-Hermitian operator is 
\begin{equation}
    \hat{\mathcal{H}}_\textrm{eff} = \hat{H} - i \left(\frac{\Gamma_0}{2}\left(\hat{\sigma}_1^+ \hat{\sigma}_1^- + \hat{\sigma}_2^+ \hat{\sigma}_2^- \right) + \kappa_{12} \left(\hat{\sigma}_1^+ \hat{\sigma}_2^- + \hat{\sigma}_2^+ \hat{\sigma}_1^-\right)\right)\, .
\end{equation}
We start by writing $\hat{\mathcal{H}}_\textrm{eff}$ in the two-spin basis of $\ket{1} = \ket{g_1 g_2}, \ket{2} = \ket{e_1 g_2} , \ket{3} = \ket{g_1 e_2}, 
\ket{4} = \ket{e_1 e_2}$ as
\begin{equation}~\label{eq:H_eef matrix form}
\mathcal{H}_\textrm{eff} = 
    \begin{bmatrix}
    0 & - \Omega_1 & - \Omega_2 & 0 \\
    - \Omega_1^* & \left(\Delta - i\frac{\Gamma_0}{2}\right) & - \left(\epsilon_{12} + i\kappa_{12}\right) & -\Omega_2 \\
    - \Omega_2^* & - \left(\epsilon_{12} + i\kappa_{12}\right) & \left(\Delta - i\frac{\Gamma_0}{2}\right) & - \Omega_1 \\
    0 & - \Omega_2^* & - \Omega_1^* & \left(2\Delta - i \Gamma_0\right)
    \end{bmatrix} \, .
\end{equation}
The effect of dipole-dipole interaction can be clarified in the weak-probe limit, i.e. $\Omega_{1,2} \approx 0$, where the following new states can be determined
\begin{eqnarray}~\label{eq:2-atom eigenstates}
    \omega_1 & = 0 \longrightarrow \ket{1} \nonumber \, , \\
    \omega_2 & = \left(\Delta - \epsilon_{12} \right) - \frac{i}{2} \left(\Gamma_0 + 2 \kappa_{12}\right) \longrightarrow \ket{+} = \frac{\ket{2} + \ket{3}}{\sqrt{2}} \nonumber \, , \\
    \omega_3 & = \left(\Delta + \epsilon_{12} \right) - \frac{i}{2} \left(\Gamma_0 - 2 \kappa_{12}\right) \longrightarrow \ket{-} = \frac{\ket{2} - \ket{3}}{\sqrt{2}} \nonumber \, ,\\
    \omega_4 & = 2\Delta - i\Gamma_0 \longrightarrow \ket{4} \, ,
\end{eqnarray}
where the real and imaginary parts of $\omega_i$ are the frequency of the normal mode and its corresponding decoherence rate, respectively.

As can be seen, the interaction removes the degeneracy of the two entangled states $\ket{\pm}$, splitting them to $\Delta \mp \epsilon_{12}$ decaying at $\Gamma_0 \pm 2\kappa_{12}$, respectively. Typically, the mode with 
the faster (slower) decay rate compared to an isolated atom is called the super (sub)-radiant state.

In general, $\mathcal{G}$ describes a non-local and retarded interaction however by limiting oneself to the near-field only, i.e. $k_0 R \ll 1$, $\mathcal{G}$ simplifies to 
\begin{equation}~\label{eq:near-field}
    \mathcal{G} \approx \frac{3 \mathbf{r} \left(\mathbf{r} \cdot \mathbf{d}_1 \right) - \mathbf{d}_1}{4\pi \epsilon_0 R^3} \cdot \mathbf{d}_2 \, .
\end{equation}
This limit is typically referred to as the \emph{coherent} interaction, where the interaction has a dispersive-only contribution ($\epsilon_{12} \ne 0$) without any dissipative part ($\kappa_{12} = 0$). 

It is straightforward to expand Green's function for small inter-atomic separations $k_0 R$ 
to find
\begin{equation}~\label{eq:dissipative expantion}
    \mathbf{d}_1 \cdot \textrm{Im(G)} \mathbf{d}_2 \simeq \frac{k_0}{6\pi} \mathbf{d}_1 \cdot \mathbf{d}_2
    -\frac{k_0^3 R^2}{60\pi} \left[2 \mathbf{d}_1  \cdot \mathbf{d}_2 -(\mathbf{d}_1 \cdot \mathbf{r}) 
    (\mathbf{d}_2 \cdot \mathbf{r})\right]\, ,
\end{equation}
where the first term is responsible for the decay rate $\Gamma_0$, and the second term accounts 
for a finite propagation time between the two emitters.

From the near-field form of Eq.~(\ref{eq:near-field}) it is evident that in an atomic ensemble, the strongest interaction arises from two dominant 
arrangements of the atoms being arranged either side-by-side (blue-colored spheres) or head-to-tail (red-colored spheres).
In the first case, $\mathbf{r}\cdot \mathbf{d}_1 = 0$, the interaction energy is positive and 
hence repulsive which leads to an increase in the energy and a blue-shift of the two-atom 
spectrum. On the other hand, for the head-to-tail configuration, the interaction energy is 
negative, meaning an attractive potential and hence a red-shifted two-atom spectrum. Besides, as 
the near-field interaction scales as $1/R^3$ (cf. Eq.~(\ref{eq:near-field})), the energy shift 
$\epsilon_{12}$ becomes proportional to the ensemble density.

\begin{figure}[h]
	\begin{subfigure}[b]{0.6\textwidth}
			\centering
			\includegraphics[width=1\textwidth]{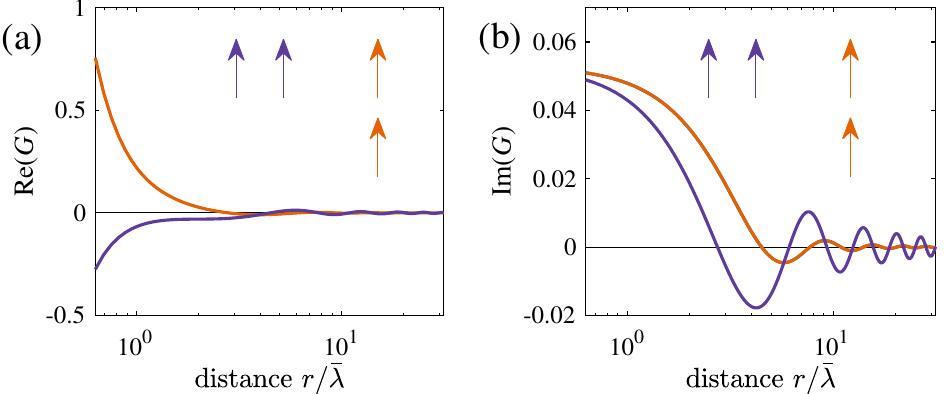}
	\end{subfigure}
	\hfill
	\begin{subfigure}[b]{0.4\textwidth}
			\centering
			\includegraphics[width=0.9\textwidth]{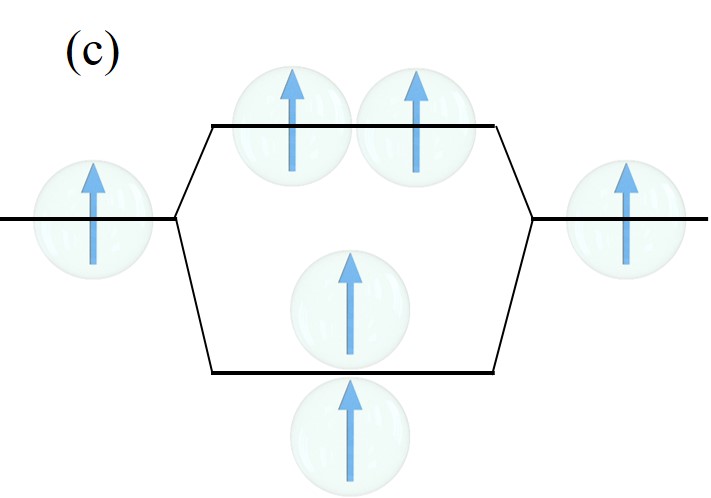}
	\end{subfigure}
	\caption{\textbf{Free-space dipole-dipole interaction of two polarized atoms.}  (a) The DDI-induced energy shift, vs. the distance between two emitters $r$ in units of the reduced wavelength $\bar{\lambda}$. (b) The DDI-induced broadening, for the same dipole orientations as in (a). (c) The energy of the two-atom case for two different orientations. In each case, configurations are depicted with arrows and higher levels indicate larger energies.}
    \label{fig:GvsR}
\end{figure}

Beyond the near-field regime, however, Green’s function oscillates with distance $R$ as a 
result of the retardation, hence the DDI-induced energy shift and the decay rates become strongly space-dependent. Figure~\ref{fig:GvsR}(a),(b) shows the real and imaginary parts 
of the light-induced interaction between two atoms as a function of their separation, 
respectively. In each panel, the side-by-side configuration case is shown in blue 
while the orange line shows the behavior of the head-to-tail configuration. Figure~\ref{fig:GvsR}(c) 
summarizes the energies of the two interacting atoms when arranged head-to-tail or side-by-side compared to the individual atom energy. 

\begin{figure}[h]
	\centering
	\includegraphics[width=0.9\textwidth]{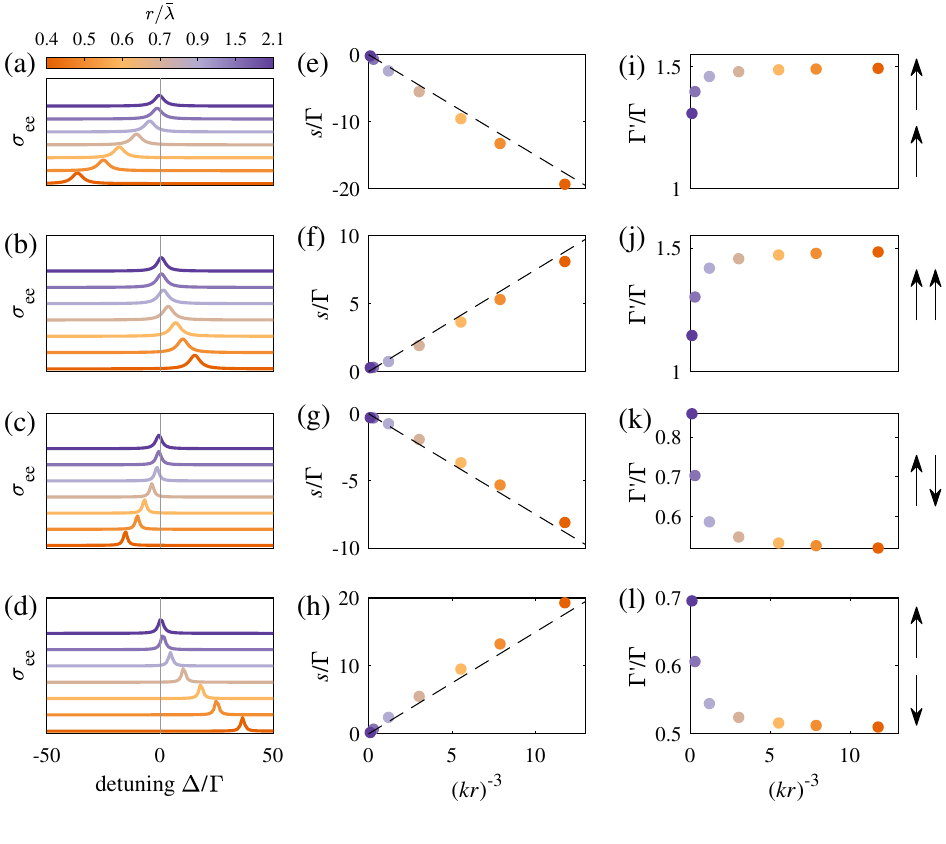}
	\caption{\textbf{DDI vs. detuning and inter-atomic separation.} (a)-(d) Waterfall plot of the excited state population $\braket{\hat{\sigma}_{ee}}$ of two stationary atoms as a function of probe detuning $\Delta$ at different inter-atomic spacings. (e)-(h) The lineshift $s$ and (i)-(l) linewidth $\Gamma'$ of the spectrum vs. normalized distance $(kr)^{-3}$ extracted from the corresponding spectrum in (a)-(d). All plots are color-coded with the colormap in the top left corner of the panel (a) that shows the inter-atomic distance $r$ in units of the reduced wavelength $\bar{\lambda}$. The rows are separated in head-to-tail (a,e,i), parallel side-by-side (b,f,j), anti-parallel side-by-side (c,g,k), and anti-parallel head-to-tail (d,h,l) configurations as depicted with arrows in each row at the right edge. Dashed lines in (e)-(h) mark simplified linear predictions $\propto r^{-3}$ for the lineshift. The driving field strength is $\Omega = 0.01\Gamma$ within the weak-probe limit.}
    \label{fig:Sfar}
\end{figure}

Figure~\ref{fig:Sfar}(a)-(d) shows the excited state population of the two fixed atoms $\braket{\hat{\sigma}_{ee}}$ vs. detuning at different 
normalized inter-atomic spacing $r$, and four possible dipole orientations shown with black 
arrows. The corresponding lineshift, shown in Fig.~\ref{fig:Sfar} (e)-(h), is extracted by 
fitting a Lorentzian to each spectrum, and the resonance is plotted as a function of the 
normalized inter-atomic spacing $(k r)^{-3}$. As one can see, the induced lineshift scales linearly with 
the density (in 3D) due to the near-field $R^{-3}$ scaling in Eq.~(\ref{eq:near-field}).

As discussed before, an attractive interaction leads to a negative shift and a repulsive 
interaction to a positive one. Further, it is straightforward to check that for parallel spins, 
i.e. $\mathbf{d}_1 \cdot \mathbf{d}_2 \ge 0$, the dipole-dipole interaction in the side-by-side 
arrangement increases the energy by $3/4 \Gamma_0$, while the head-to-tail configuration reduces 
the energy by $3/2 \Gamma_0$. The estimated lineshift from the near-field approximation is plotted as a dashed line in Fig.~\ref{fig:Sfar}(e)-(h), which is close to the exact solution obtained from the full Green’s function, shown in circles. In both 
cases, the attractive shift of the head-to-tail configuration is twice as strong as the repulsive 
shift of the side-by-side one. This simple near-field interaction picture will be useful when we 
investigate dipolar interactions in different dimensions in the next section. 

Figure~\ref{fig:Sfar} (i)-(l) illustrates the dissipative effect of the dipole-dipole interaction 
in modifying the decay rates of the collective states. For both optically-active cases in 
Fig.~\ref{fig:Sfar}(i) and (j), the effective decay rate of the entangled states $\Gamma'$ is 
larger than the free space lifetime, hence it is a super-radiant mode. On the other hand, for the 
optically dark states shown in Fig.~\ref{fig:Sfar}(k) and (l), the effective decay rates of the entangled states are smaller than the isolated atom implying a longer-lived and hence a sub-radiant 
state.

\subsection{Dipole-dipole interaction in a thermal cloud in lower dimensions}~\label{subsec: DDI at low-D}
To model the experiments involving thermal atoms subject to motional dephasing and the 
Doppler effect, as well as studies on cooperative effects in solid-state systems with spectral 
wandering, a detailed investigation of the inhomogeneous effects is required. In such cases, it 
is no longer sufficient to consider the fixed atoms as done so far. Instead, one has to 
take into account the random velocities and trajectories of individual atoms as well as a time-
varying mutual interaction. The resulting Doppler and transit time effects play crucial roles in 
the optical properties of the ensemble and in modifying the static picture. To that end, we 
developed a Monte-Carlo algorithm to evolve the density matrix of the atomic ensemble in time, 
using the quantum master equation derived in Eq.~(\ref{eq:atomic reduced density matrix2}). We 
take into account the atoms in the ensemble with random velocities and trajectories that can be 
updated at each time step to fully capture the motional effects such as Doppler and finite-time 
atom-field interactions. 

\begin{figure}[h]
\centering
	\includegraphics[width=0.9\textwidth]{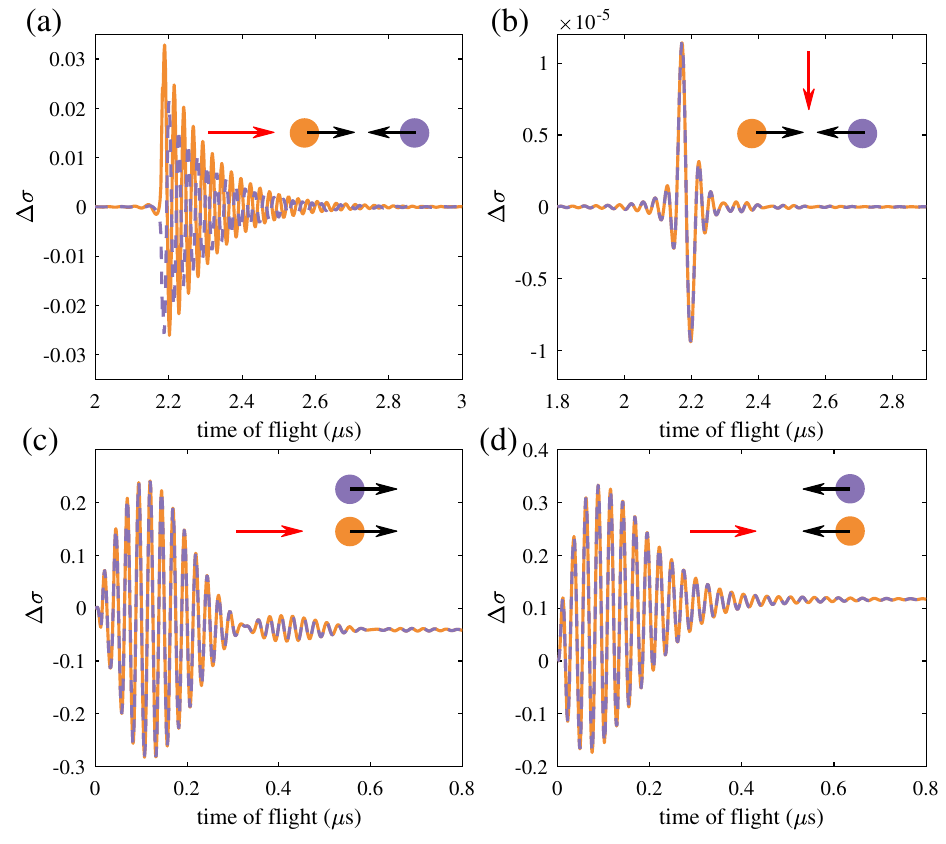}
	\caption{\textbf{DDI between two flying atoms.} Dynamic dipole-dipole interaction of two flying atoms. The difference in the excited state population $\Delta\sigma$ between two interacting atoms and two non-interacting ones is plotted as a function of time-of-flight. The orange and purple curves correspond to individual atoms' quantities. Four different probe laser (red arrow) configurations are considered: (a) counter-propagating atoms with (anti-) parallel probe, (b) counter-propagating atoms and orthogonal probe, (c) co-propagating atoms with a parallel probe, and (d) co-propagating atoms with an anti-parallel probe. The laser is linearly polarized with an out-of-plane field. The absolute velocities of all atoms are equal in each scenario. The driving field is weak $\Omega = 0.01\Gamma$ and the probe is resonant, i.e. $\Delta = 0$.}
    \label{fig:2atomfly}
\end{figure}

Figure~\ref{fig:2atomfly} shows some important possible dipolar collisions with moving atoms in a typical thermal cloud. In all panels, we plot 
the difference in the excited state population between the interacting atoms 
$\braket{\hat{\sigma}_{ee}^\textrm{int}}$ and the non-interacting case 
$\braket{\hat{\sigma}_{ee}^\textrm{0}}$
with $\Delta \sigma = \braket{\hat{\sigma}_{ee}^\textrm{int}} - \braket{\hat{\sigma}_{ee}^0}$. 
This way, we isolate the dynamics of the atomic states in the driving field from the interaction effects. 

In Fig.~\ref{fig:2atomfly} (a), two atoms start at $t=0$ at a distance $R$ and fly towards each 
other with identical velocities $v_1 = -v_2$. The atoms are shown as orange and purple circles, 
while the direction of motion of the atoms and the laser direction are depicted as black and red 
arrows, respectively. At $t \approx 2.2 ~\mu s$, the atoms meet, an event that leads to a change 
in $\Delta \sigma$ as a result of the interaction. Further, as the dipolar forces are weak, 
they have negligible effects on the velocity of the fast-moving thermal atoms and hence the motion of atoms can be treated classically. The resulting 
oscillations of the excited state population quickly vanish as the distance between both atoms 
increases after the collision. Note that, the evolution is not identical for both atoms as 
they experience different Doppler shifts concerning the probe laser. 

When the probe direction is perpendicular to the atoms' trajectories as shown in 
Fig.~\ref{fig:2atomfly}(b), the scenario for both atoms is symmetric, and hence both atoms evolve 
equally due to the driving laser as well as the mutual dipolar interactions. Unlike case (a) 
here, as soon as both atoms interact, they become detuned to the probe laser which 
leads to an overall smaller effect.

In Fig.~\ref{fig:2atomfly} (c) both atoms move in the side-by-side configuration and start at a 
distance of $R = \lambda/20$ with the same velocity $v_1 = v_2$. Due to their initial configuration, they interact from the start $t = 0$ and evolve towards a new steady state based on their identical detuning relative to the 
probe. As both atoms are in the near-field of one another at all times, they experience a strong 
repulsive interaction and hence a dipolar-induced blueshift. However, due to the relative motion 
of the atoms and the laser propagation, the driving field is Doppler shifted to lower energies, 
i.e. red-shifted. As a result, both atoms are shifted out of resonance which reduces their 
interaction. This is observable as a reduced oscillation. As their interaction decreases, the 
interaction-induced blueshift is reduced as well and hence their detuning decreases which consequently leads to their stronger polarization. This behavior is noticeable as a 
revived population in Fig.~\ref{fig:2atomfly}(c) at $t\approx 0.35$. A similar scenario is depicted in 
Fig.~\ref{fig:2atomfly}(d), however, as both atoms travel toward the laser, the blue-detuned laser will be in resonance with the blueshifted atoms. Due to this resonant driving 
field and hence increased polarization, both atoms interact strongly with one another as can be seen in the amplitude of the 
oscillations compared to Fig.~\ref{fig:2atomfly}(c).
All other interactions in an ensemble resulting from arbitrary trajectories of two atoms can be 
thought of in these four cases summarized in Fig.~\ref{fig:2atomfly} with the only differences 
being the randomized orientations of the atoms and a larger coordination number.

\begin{figure}[h]
	\centering
	\includegraphics[width=0.8\textwidth]{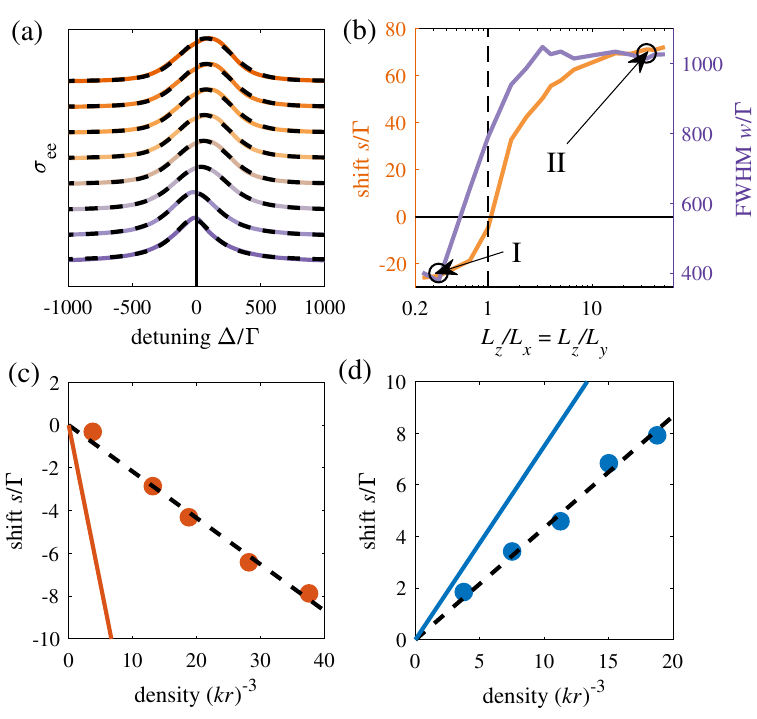}
	\caption{\textbf{Tunability of DDI with the ensemble dimensionality.} Continuous transition of dipolar interactions from 2D to 1D without Purcell enhancement. (a) Waterfall plot of the excited state spectrum for different aspect ratios between transversal and longitudinal dimensions. The ratio $L_z/L_x = L_z/L_y$ increases from bottom to top. Dashed lines show Voigt fits into the spectrum. (b) Lineshift and linewidth extracted from the Voigt fit as a function of transverse to longitudinal aspect ratio. Region I and Region II  mark the regime where attractive and repulsive interactions dominate the spectrum, respectively. The density is kept constant for each aspect ratio. The transversal dimensions are $L_x = L_y = 0.3\lambda$. (c) Lineshift as a function of density for a 2D (I) and a 1D (II) probe volume referring to the markers in (b). Points represent simulated lineshifts, dashed lines are linear fits and solid lines show the expected lineshift for a simple picture of a static dipole ensemble in the near field with the associated dipole orientation. The driving field strength is $\Omega = 0.01\Gamma$.}
    \label{fig:1Dto2D}
\end{figure}

The previous results imply that the effect of dipolar interactions can be modified noticeably by changing the atomic orientations or the coordination numbers, so it is interesting to explore if the well-known cooperative effects can be modified via the ensemble \emph{geometry or dimensionality}. From the two-atom discussions, it is evident that the ensemble can be confined such that certain 
dipole configurations are favored, e.g. the head-to-tail configuration dominates the ensemble when confined to 2D. To study this more systematically, we start with a 2D ensemble with $L_x = 
L_y = 0.3\lambda$ and $L_z = 0.05\lambda$ and continuously increase the volume by increasing $L_z$ such that the initial 2D geometry is changed to a pseudo-1D rectangular "tube". 

The corresponding excited state population vs. detuning for various lengths and the energy shift extracted from these spectra as a function of the tube length transitioning from 2D to 1D is 
depicted in Fig.~\ref{fig:1Dto2D}(a) and (b), respectively. Note that the linewidth is a result of the Doppler and transit time broadening, due to the small transversal dimensions $L_{x,y}$, as well as the frequent boundary collisions. The Doppler detuning emerges from the velocity component along the z-axis (parallel to the laser) 
which did not exist in the pure 2D case. The Doppler broadening continuously vanishes with decreasing $L_z$. The atoms with a large $v_z$ along the laser immediately collide with the wall and only scatter a small amount of the probe light, while atoms with small $v_z$ interact longer with the probe field and hence 
contribute to the signal more. This gradual contribution of the dephasing with increasing $L_z$ leads to the broadening depicted in Fig.~\ref{fig:1Dto2D}(b). For $L_z \ge \lambda$ the broadening 
remains almost unchanged as the volume is large enough to fully resolve the dephasing along the $z$-axis and an equilibrium condition for the transversal collisions, along the $x$- and $y$-axis, is reached. The lineshift depicted in panel (b) is extracted from a Voigt fit to the simulated data in Fig.~\ref{fig:1Dto2D}(a), shown as dashed lines in the same panel.

One can see from the orange line in Fig.~\ref{fig:1Dto2D}(b) that the interaction varies 
from an attractive to a repulsive corresponding to red and blue shifts, for $L_z \le L_x$ and 
$L_z \ge L_x$, respectively. The former is a quasi-2D scenario where the head-to-tail 
configurations of the dipoles are dominant which leads to a redshift of the averaged spectrum. In 
contrast, the latter corresponds to an elongated pseudo-1D cloud and accommodates more side-by-side atom pairs with repulsive interactions which leads to a blueshift, ultimately. At $L_x = 
L_y = L_z$, denoted by a vertical dashed line in Fig.~\ref{fig:1Dto2D}(b), the near-field dipole 
interactions from various arrangements cancel out each other and the collective shift vanishes. 
Hence, a rearrangement of dipoles using different confinements can switch the collective behavior from attractive to repulsive without manipulating each atom. Note that both, red-shift as well as blue-shift, saturate at $L_z/L_x \le 0.1$ and $L_x/L_z \ge 10$, 
respectively, implying that at least an aspect ratio of $\approx 10$ between transversal and 
longitudinal confinement is required to render the system to 2D or 1D.

Figure~\ref{fig:1Dto2D}(c) and (d) show the lineshift for different densities at simulation conditions marked with "I" and "II" in panel (b). Both shifts scale linearly with the normalized 
density $(k_0 R)^{-3}$ as shown with linear fits (dashed line) to the simulated data (points). However, when compared to the near-field approximations (solid lines in Fig.~\ref{fig:1Dto2D}(c),(d)), the collective effect is noticeably weaker. This can be attributed to the perturbed repulsive interactions resulting in a blueshift in (II) by attractive interactions as the transversal dimensions are still finite such that occasionally atom pairs can be found in the head-to-tail 
configuration. 

After averaging, the excited state population over all atoms and time steps the effect of confinements along the $x$- and $y$-axis are noticeable as a reduction of the expected blueshift 
in a 1D ensemble. The same is true for the attractive interactions in (I). Overall, Fig.~\ref{fig:1Dto2D} shows that moving atoms confined in different free-space geometries do not 
behave similarly to the simple static two-atom cases explained in the
previous section and one cannot extend the simple near-field interpretation as described with Eq.~(\ref{eq:near-field}) to the free-space ensemble behavior. Also, the aspect ratio regarding 
transversal and longitudinal components plays a crucial role in the energy shift of the spectrum.

\section{Experiments}~\label{sec: experiment}
Although collective light shifts caused by light-induced dipole-dipole interaction were predicted 50 years ago in a seminal paper by R.~Friedberg \textit{et al.}~\cite{Friedberg72}, it took a long time 
until the first experimental evidence of such a shift was provided. The actual shift depends on various parameters including the shape of the sample, the density, the level structures, the type of quantum 
emitters, and whether hot or ultracold atoms are used. 
As a side remark, the dipole-dipole shift is often termed the cooperative Lamb shift. However, this effect is not related to the fluctuations of the electromagnetic field, i.e. the origin of the Lamb 
shift derives from QED (cf.~\ref{app:CP}). Although R.~Friedberg \textit{et al.}, who initially coined this term, revisited their wording later~\cite{Friedberg2022}, the term 'Lamb shift'
is still meaningful in a historical context~\cite{Scully2010}.

The first attempt to measure the collective response was conducted in an optically-excited thermal xenon gas in 1990~\cite{Garrett1990}. There, a three-photon excitation scheme in a macroscopic cell 
filled with xenon at high densities and large laser beams compared to
wavelengths were used. The excitation scheme was chosen to avoid radiation trapping and incoherent multiple scattering, which can also lead to line broadenings and shifts~\cite{Chevrollier2012, 
Molisch1998}. 

Further, laser beams were aligned at multiple angles to create 3D excitation volumes. Predominantly, the Lorentz-Lorenz shift, \emph{aka} the Clausius–Mossotti shift, is anticipated to be the primary contributor to the observed line shift, typically manifesting as a redshift. Additionally, there are minor corrections attributable to collective effects, which are highly sensitive to the geometry of the excitation volume. However, these corrections are presumed to be minimal in a 3D sample, as the net effect of all dipole-dipole interactions (DDIs) is expected to average out to zero. Nonetheless, the obtained data shows an overall blueshift for most of the configurations. Compared to later experiments carried out in rubidium vapors as will be discussed in what follows, the shifts in xenon are roughly an order of magnitude larger. 

In~\cite{Friedberg1989}, the authors have analyzed the shifts in various geometrical arrangements, including 3D configurations, identifying three key components of the lineshift: 1) the Coulomb term, 2) the cooperative Lamb shift (CLS), and 3) the resonant collision shift.

For larger volumes, the Coulomb term also referred to as the Lorentz-Lorenz shift, is quantified as \(\Delta_{\textrm{LL}}=-\frac{N d^2}{3\epsilon_0 \hbar}\), where \(N\) represents the ensemble density and \(d\) the transition dipole moment. The CLS conceptualized as a virtual photon exchange between two atoms, introduces a correction in 3D, expressed as \(\Delta_{\textrm{CLS}} = \frac{\Delta_{\textrm{LL}}}{4}\). The final component, the resonant collision shift, arises from transient correlations in dipole orientations during close encounters between atoms, involving a permanent exchange of excitation.

From the available information, it is hard to delimit the effect of multiple scattering. Even if such parasitic effects can be neglected, other effects such as excitation above 
saturation, transient interaction, Doppler selection, optical pumping, polarization, and laser linewidth may have influenced the experimental outcomes. Therefore, a re-investigation of these measurements in a 
more controlled experiment using today's technology and modern narrow-band CW lasers will be beneficial.

It is worth emphasizing that, our approach and the numerical results incorporate all these factors, with the limitation that correlations beyond two-body interactions were not considered. Our simulations also indicate a shift towards 3D characteristics (cf. Fig.~\ref{fig:1Dto2D}). However, due to the limited size of the systems that we can simulate, it remains unclear how significant the impact of finite-size effects is in this context.

In another experiment, X-rays were used to excite a solid iron sample inside a low-Q 
cavity~\cite{Roehlsberger2010}. They use a 14.4 keV excitation source, i.e. a wavelength of 86 pm, which 
is smaller than the iron lattice constant of 287~pm. The advantage of this experiment is the absence of 
atom-atom interactions, which simplifies the description. The collective response of the system was 
studied in terms of Dicke superradiance accompanied by a collective shift~\cite{Gross2002} and measured 
in the spectra of the superradiant burst. Further, the fluorescence lineshift showed a scaling with the 
number of iron atoms. This is not the typical 
density scaling as the density of the iron samples is constant. The larger number of atoms in a bigger 
iron sample $N$ increased the number of atoms participating in a Dicke state, leading to an 
$N$-dependent lineshift and the superradiant time scale. The observed shifts 
are about 5 orders of magnitude smaller than those observed in rubidium (when compared via particle 
density), but of course, the x-ray wavelength is about 4 orders of magnitude smaller. This experiment is also hard to compare to Friedberg's original proposal~\cite{Friedberg72} since the collective effects are not due to light-induced DDIs. The distance between the quantum emitters is larger than the involved wavelength, a situation closer to the canonical Dicke model. Besides this, the experiment has been carried out in a cavity and in a transient regime, both of 
which are not present in the original Friedberg paper, either. Nonetheless, the theoretical model applied by the authors describes the observations rather well.

As geometry plays a crucial role in modifying the collective effects, it is desirable to obtain good control over the actual arrangement of the quantum emitters and the excitation volume. The emitters should be as simple as possible, ideally, two-level systems, to simplify the description of the excitation dynamics and to reduce the complexity 
of any theoretical or numerical simulation. Alkali atoms with their lowest lying S to P transitions in the VIS/NIR domain are ideal candidates for this. The strength of the light-induced dipole-dipole interaction becomes significant when the distance between the atoms gets smaller than the involved wavelengths. This can easily be achieved with thermal gases of alkali atoms, \emph{e.g.} potassium, rubidium, and cesium, by heating the metal to 500-700~K. The drawback of such dense gases is their large optical depth and hence the very low signal-to-noise ratio. To study light-induced dipole-
dipole interactions it is necessary to stay below the saturation intensity, i.e. a few $mW/cm^2$. With densities similar to a BEC (e.g. at around 150 Celsius in a rubidium vapor) one obtains optical depths 
above 1 at a $1\mu m$ thickness~\cite{Baluktsian2010}. The solution is to use vapor cells with an extremely narrow spacing, ideally with a wedge-shaped spacing to adjust the thickness of the sample to 
thicknesses below $1~\mu m$. 

With such wedged cells, two experiments have been performed in rubidium~\cite{Keaveney2012} and potassium~\cite{Peyrot2018}. The atoms have been excited in cell lengths ranging from $(10~nm - \mu m)$ 
with laser beam diameters larger than the cell thickness and the excitation wavelength. The result is a 2D pancake-shaped atomic cloud where the expected cooperative collective shift is~\cite{Friedberg72}   
\begin{equation}\label{LL}
	\Delta_\textrm{CLS} =\Delta_\textrm{LL} - \frac{3}{4}\Delta_\textrm{LL} \left(1-\frac{\sin 2kL}{2kL} \right)\, ,
\end{equation}  
where $L$ is the length of the vapor cell. 

For a very short vapor slab, the system is effectively two-dimensional. For high enough densities, the interaction is dominated by the non-retarded light-induced DDI in a plane, where the polarization lies in the plane of the atoms, and the attractive head-to-tail contribution of the dipole-dipole interaction dominates over the repulsive side-by-side contribution resulting in a net red-shift~(cf. Sec.~\ref{sec: 
interacting cloud} for detailed discussions). When extending the sample to $L=\lambda/4$, the retardation of the propagating laser leads to a $\pi$ phase shift. The dipoles are now oriented in the 
opposite direction and the interaction results in a blueshift. Therefore, one expects an oscillation of $\Delta_\textrm{CLS}$ as a function of the cell thickness which is consistent with the experimental observations. Further, the theory developed in~\cite{Friedberg72} matches well with the experimental data although it does not consider the full quantum correlations between the atoms. A remaining oscillation of the data as a function of cell length can be explained by the effective refractive index of the atomic slab itself. In addition, there is a cavity effect induced by the cell walls as theoretically described in~\cite{Dobbertin2020}.

To avoid the undesired cavity effect of the wedged cells, one could employ the Light-Induced Atomic Desorption (LIAD) effect to desorb atoms from the cell surface via a pulsed laser. Using a short pulse, 
compared to the motion of the released atoms, for a very short time a 2D ensemble of atoms can be created. At high laser intensities, the density of this 2D sample can be large and the light-induced DDI effects can be studied. This has been recently observed in~\cite{Christaller2022}, where the scaling with the transition strength has been studied, by confirming the quadratic scaling of the lineshift between the D$_1$ (795 nm) and D$_2$ (780 nm) transitions in rubidium. 

Another potential candidate to measure collective lamb shifts in a 2D geometry is based on the total reflection of light from the glass wall of a vapor cell, where a very short-range evanescent field can be established inside the atomic vapor. This technique was used in a 
pioneering experiment in 1991 to observe the Lorentz-Lorenz shift in potassium vapor~\cite{Maki1991}. However, a clear signature of cooperative Lamb shifts has not been reported in similar experiments, so far.

Finally, there exists one experimental result on a cooperative line shift in a 1D atomic ensemble~\cite{Skljarow2022}. The geometry, in this case, is set by the light mode confined in a slot waveguide. Since the slot is only 50 nm which is much smaller compared to the 1529 nm, the wavelength of the 5P-4D transition in rubidium used in that work, the atoms entering the light mode inside the waveguide are effectively arranged in one dimension. The polarization is perpendicular to the atomic chain, and hence the induced dipoles are parallel to each other. For close-by atoms, the dipole-dipole interaction is repulsive and the line shift goes predominantly to the blue. An interesting aspect of having an optical structure close to the atoms is an enhanced Purcell factor favoring scattering into the waveguide mode~\cite{Asenjo-Garcia2017}. 
This not only enhances the DDI along the waveguide but also changes the 1/r$^3$ scaling in free space to an all-to-all interaction. To describe the observed line shift, the states of the atoms in a non-uniform field were determined using Bloch equations while the full motion of the atoms was included. In addition, the Casimir-Polder shifts of the atomic resonances due to the waveguide as well as a pairwise light-induced DDI have been included. This method, as detailed in Sec.~\ref{sec: interacting cloud}, led to an excellent agreement with the experimental data~\cite{Skljarow2022}. 

Over the last 33 years, there have been so far only 5 experimental results on the light-induced cooperative phenomenon in atoms. The sixth result is the thin iron foil experiment in a low-Q cavity, as 
described above. In thermal gases, the short coherence time allows a simplified theoretical treatment that neglects all higher-order quantum correlations, except the pair correlations between two atoms. For ultracold atoms, however, the behavior cannot be fully captured unless higher-order correlations are included. 

First experiments have been carried out in one-dimensional systems~\cite{Glicenstein2020, Maiwoeger2022}, but more experiments are on the way. The potential to arrange ultracold atoms in optical tweezers in arbitrary configurations has inspired a lot of 
theoretical studies during the last five years. One candidate to implement many of these ideas relies on the low-lying clock transitions in the alkaline earth elements~\cite{Olmos2013}. They are especially suited to study coherent collective and cooperative effects in ultracold atoms due to their long wavelength and the possibility of their side-band cooling to the ground state inside the tweezers.

In parallel, there are attempts to use ensembles of solid-state quantum emitters to study collective effects~\cite{Pak2022, Davis2023}. Solid-state emitters have several advantages compared to 
atoms as they are fixed in space, they can be closely packed to high densities, their excitation and emission wavelength are flexible, and they can be easily integrated with nano-photonic structures. Their main drawback is the dispersion of absorption/emission wavelengths among \emph{identical emitters}, which makes it hard to observe collective effects. Also, the coupling to the environment, e.g. phonons, spin noise, and charge noise, leads to additional dephasing and shorter coherence times. We are not aware of any experiment observing light-induced interactions with many solid-state emitters thus far, except for a work observing a large collective Lamb shift of two coupled superconducting ``atoms''~\cite{Wen2019}. The key is the tunability of the emitters by external electric fields or via local strain of the supporting solid-state matrix. This has been achieved for optically active molecules frozen onto a surface at liquid helium temperature. Here two molecules were brought into resonance by electric field tuning~\cite{Hettich2002}. Also, three strain-tuned quantum dots coupled via a photonic waveguide have been enabled to superradiate collectively~\cite{Grim2019}. How these methods can lead to a scalable approach has to be investigated in the future. 

What are the open questions an experiment can address? There is still only limited knowledge regarding the influence of geometry and dimensionality on the collective line shifts. Also, absorption 
and fluorescence have different properties, and the connection between super-/subradiance with the various sources of line shifts is not entirely understood. The role of coherence and quantum correlations is not very significant in thermal gases, but with longer coherence times in ultracold atoms, and ensembles of fixed solid-state emitters they become more relevant. Finally, one has to distinguish between the steady-state features and the transient responses. Both regimes are 
relevant to exploit dense ensembles of quantum emitters for quantum technologies, e.g. quantum sensors, single-photon storage, or light-driven quantum computers and simulators.

\section{Conclusion}~\label{sec: conclusion}
Our theoretical framework for examining thermal vapors within macroscopic environments marks an initial step towards exploiting the considerable adjustability afforded by this platform. However, it raises several pressing queries that require further exploration.

Firstly, there is a need to extend our coupled dipole model to accommodate large atomic densities analogous to those encountered in experimental setups. Moreover, it is imperative to include the temporal dynamics, accounting for atomic motion. This temporal consideration becomes pivotal in addressing nonlocal effects, variations in Doppler shifts, atomic collisions, and other dynamic phenomena. However, the principal impediment to this endeavor is the substantial numerical complexity involved in seeking exact solutions to the coupled dipole model. Recent advances, exemplified by the 
renormalization group approach, offer auspicious avenues to approximate solutions and potentially alleviate the computational demands~\cite{Grava2022}. This becomes particularly relevant if avenues for deriving (semi-)analytical solutions can be explored. An initial step entails investigating the feasibility of 
extending this approach to encompass the precise three-dimensional Green’s tensor associated with macroscopic geometries and stochastically include the Doppler effects.

Secondly, our developed coupled-dipole model, originally applied to thermal, exhibits potential for broader applicability. It can be extended to scrutinize atom-atom interactions across a wider spectrum of macroscopic geometries. While extensive investigations into atom-atom interactions have occurred in one- and two-dimensional photonic crystals, the interaction between hot or cold atoms and diverse structures, such as microresonators, nanofibers, hollow-core fibers, 
superconducting chips, and atomic cladding waveguides, remain unexplored. 

Lastly, it is important to extend the analysis to Rydberg atoms. In numerous experiments, atom-atom interactions were of minimal concern due to low atomic densities. However, this scenario undergoes a transformation when atoms are excited to higher energy levels, particularly in Rydberg states. Transitions between Rydberg levels feature significantly smaller wavenumbers, necessitating fewer atoms to reach the critical density of $1/k^3$. Consequently, the coupled-dipole model needs to be extended in two critical aspects. Firstly, it should account for coherences between multiple atomic levels, given that Rydberg atoms are typically produced through a two-step excitation process. Secondly, the treatment of atom-wall 
and atom-atom interactions should adopt a self-consistent approach. Rydberg atoms are characterized by potent interactions and minute energy gaps between adjacent states, suggesting that the impact of 
interactions may no longer be perturbatively small. While many existing studies have addressed these aspects separately, a unified model capable of addressing dense thermal vapors within arbitrary 
geometries remains elusive.

Ultimately, once we have comprehensively described all pertinent aspects of a system, the macroscopic quantum electrodynamics (mQED) formalism can showcase its full potential. A macroscopic environment, 
represented by the Green’s tensor, can be tailored to induce specific system behaviors. Over the long term, this inverse design approach holds the potential to enhance existing applications, including vapor-based single-photon sources and quantum memories, while also opening the door to entirely novel yet-to-come applications.

\section*{Acknowledgments}
The authors would like to acknowledge stimulating and insightful discussions with H. Dobbertin and C. S. Adams. HA acknowledges financial support from the Eliteprogramm of Baden-W\"urttemberg Stiftung, DFG (GRK2642 "Photonic Quantum Engineers") through the Mercator Fellowship, the Purdue University Startup fund, the financial support from the Industry-University Cooperative Research Center Program at the US National Science Foundation under Grant No. 2224960, and the Air Force Office of Scientific Research under award number FA9550-23-1-0489. We gratefully acknowledge Deutsche Forschungsgemeinschaft (DFG) through the Priority Programme 1929 'Giant Interaction in Rydberg systems (GiRyd)'.

\section*{References}
\bibliographystyle{iopart-num}
\bibliography{Ref}

\newpage
\appendix
\section{Casimir-Polder and Lamb Shift}~\label{app:CP}
In Eq.~\ref{eq:EoM coupled polarization} of the main text we showed how one can obtain the well-known interacting spin model after several assumptions about the relative frequencies including the coherent drive frequency as well as the resonance frequency of the emitters. In particular, when the two-level approximation is valid the coupling of far-detuned transitions to the laser can be ignored in the dipole-dipole interaction. However, the presence of these off-resonant excitations and the photon exchange leads to additional collective corrections in terms of a frequency shift and a broadening. 

In this Appendix, we present the detailed expressions for the collective off-resonant frequency shift, i.e. $\omega^{m^\textrm{CP}}$, and the collective decay due to the off-resonant transitions, i.e. $\Gamma^m$. Following the general expression of the atomic reduced density matrix and summing over all other transitions, we will get a state-dependent frequency shift as
\begin{eqnarray}~\label{eq: collective CP shift}
    \omega_{A_{\nu \delta}}^{m^\textrm{CP}} = & -\frac{\mu_0 \omega^2}{\hbar} \sum_{k,\kappa} \mathbf{d}_{A_{\nu \kappa}}^{mk} \cdot \{[\Theta(\omega_{A_,mk})\left(n(\omega_{A,mk}) + 1)\right) - \Theta(\omega_{A,km}) n(\omega_{A,km})]\nonumber \\
    & \textrm{Re} (G(\mathbf{R}_A, \mathbf{R}_A, \omega_{A,mk})) - \frac{2k_B T}{\hbar} \sum_{j = 0}^\infty \frac{\omega_{A,mk} G(\mathbf{R}_A, \mathbf{R}_A, i\xi_j)}{\xi_j^2 + \omega_{A,mk}^2} \} \cdot \mathbf{d}_{A_{\kappa \delta}}^{km} \nonumber \, ,   \\
\end{eqnarray}
as well as an additional broadening as
\begin{eqnarray}~\label{eq: collective decay}
    \Gamma_{A_{\nu \delta}}^m = & \frac{2 \mu_0 \omega^2}{\hbar} \sum_{k,\kappa} \mathbf{d}_{A_{\nu \kappa}}^{mk} \cdot \{[\Theta(\omega_{A_,mk})\left(n(\omega_{A,mk}) + 1)\right) + \Theta(\omega_{A,km}) n(\omega_{A,km})]  \nonumber \\ & \textrm{Im} (G(\mathbf{R}_A, \mathbf{R}_A, \omega_{A,mk})) \} \cdot \mathbf{d}_{A_{\kappa \delta}}^{km} \nonumber \, . \\
\end{eqnarray}
Here, $\xi_j = \frac{2\pi k_B T}{\hbar} j$ denotes the Matsubara frequencies, originating from the poles of the Bose-Einstein distribution on the imaginary axis~\cite{Scheel2008}. Remarkably, the finite temperature only changes the shift and broadening through the density of the fluctuating photons, but does not contribute to the resonant dipole-dipole interaction. As a limiting case, the broadening contains the free-space contribution at zero temperature. Furthermore, it includes the modifications of the local density of states due to the presence of any objects through the corresponding Green's function $G$. This consequently leads to a modified decay rate as predicted by Purcell~\cite{Purcell1995, Purcell1964}.

To interpret the results of the collective shift and decay let us focus on the energy levels $m,k$. The expression $\mathbf{d}_{A_{nk}} \cdot \textrm{Re}(G(\mathbf{R}_A, \mathbf{R}_A, \omega_{A,mk}) \cdot \mathbf{d}_{A_{nk}}$ describes the following phenomenon: atom $A$ undergoes the $m \rightarrow k $ transition using a photon with frequency $\omega_{A_{mk}}$ that travels from the atoms to the environment and comes back to the atom where it is reabsorbed. The transitions are either triggered by the stimulated emission and absorption of a thermal photon ($n(\omega_{A_{mk}})$) or by the stimulated emission due to vacuum fluctuations (term 1).

The free-space part of Green’s tensor formally gives an infinite result. But it can be renormalized and then gives the Lamb shift~\cite{Buhmann2008}, which is complemented by an AC-Stark shift at the finite temperature. Such a shift in the presence of some macroscopic objects and not in the free space is the so-called Casimir-Polder shift. The free space Lamb shift is naturally independent of the macroscopic environment and the position of the atoms.

\end{document}